\documentclass[pra,aps,showpacs,twocolumn,superscriptaddress]{revtex4}
\usepackage{amsmath,amssymb,amsthm}
\usepackage{graphicx,dcolumn,bm}
\usepackage{color}
\usepackage{tabularx}
\usepackage[colorlinks=true,citecolor=blue,hyperfootnotes=false]{hyperref}

\newcommand{\element}[3]{\makebox[0cm][r]{\tiny $^{^{#3}\!}$}\makebox[0.03cm][r]{\tiny $_{#2}$}#1}
\newcommand{\PrTerm}{$^4 I^o_{9/2}$}
\newcommand{\MgIon}{Mg$^+$}
\newcommand{\PrComponent}{\mbox{$\lambda$~=~1118.5397(4)~nm}}
\newcommand{\PrComponentQ}{\mbox{$\lambda / 4$~=~279.6349(1)~nm}}
\newcommand{\MgLineSH}{\mbox{$\lambda_{Mg^{+}}^{4^{th}SH}$~=~1118.540~nm}}
\newcommand{\MgLine}{\mbox{$\lambda_{Mg^{+}}$~=~279.635~nm}}
\newcommand{\MgIons}{\mbox{$^{24}$Mg$^+$, $^{25}$Mg$^+$, and $^{26}$Mg$^+$}}

\newcommand{\Dtwo}{D$_2$}

\newcommand{\MgIonLine}{\mbox{$3s \ ^2S_{1/2} - 3p \ ^2P_{3/2}$}}
\newcommand{\PrLockLine}{\mbox{$E=16\,502.616_{7/2} \ \mathrm{cm^{-1}} \rightarrow E'=\textit{25\,442.742}^{\,o}_{9/2} \ \mathrm{cm^{-1}}$}}

\begin{document}

\title{Active laser frequency stabilization using neutral praseodymium (Pr)}

\author{S. Oppel}
\affiliation{Institut f\"ur Optik, Information und Photonik, Universit\"at Erlangen-N\"urnberg, Erwin-Rommel-Stra{\ss}e 1, 91058 Erlangen, Germany}
\email{steffen.oppel@physik.uni-erlangen.de}
\homepage{http://www.ioip.mpg.de/jvz/}

\author{G. H. Guth\"ohrlein}
\affiliation{Fachbereich Elektrotechnik, Lasertechnik und Werkstofftechnik, Helmut-Schmidt-Universit\"at, Universit\"at der Bundeswehr Hamburg, Holstenhofweg 85, 22043 Hamburg, Germany}

\author{W. Kaenders}
\affiliation{Toptica Photonics AG, Lochhamer Schlag 19, 82166 Gr\"afelfing (M\"unchen), Germany}

\author{J. von Zanthier}
\affiliation{Institut f\"ur Optik, Information und Photonik, Universit\"at Erlangen-N\"urnberg, Erwin-Rommel-Stra{\ss}e 1, 91058 Erlangen, Germany}

\date{\today}

\begin{abstract}
We present a new possibility for the active frequency stabilization of a laser using transitions in neutral praseodymium. Because of its five outer electrons, this element shows a high density of energy levels leading to an extremely line-rich excitation spectrum with more than \mbox{25$\,$000} known spectral lines ranging from the UV to the infrared. We demonstrate the active frequency stabilization of a diode laser on several pra\-seo\-dym\-i\-um lines between 1105 and 1123~nm. The excitation signals were recorded in a hollow cathode lamp and observed via laser-induced fluorescence. These signals are strong enough to lock the diode laser onto most of the lines by using standard laser locking techniques. In this way, the frequency drifts of the unlocked laser of more than 30~MHz/h were eliminated and the laser frequency stabilized to within 1.4(1)~MHz for averaging times $>0.2$~s. Frequency quadrupling the stabilized diode laser can produce \mbox{frequency-stable} \mbox{UV-light} in the range from 276~to~281~nm. In particular, using a strong hyperfine component of the praseodymium excitation line \PrLockLine~at \PrComponent~makes it possible - after frequency qua\-dru\-pli\-ca\-tion - to produce laser radiation at \PrComponentQ, which can be used to excite the \Dtwo~line in \MgIon.
\end{abstract}

\maketitle 

\section{Introduction}
\label{intro}

Lasers are widely used in many fields of applied and fundamental research. However, for some applications, a serious drawback is their residual
frequency instability. Free-running lasers without any measures of active or passive frequency stabilization exhibit
\mbox{short-term} frequency fluctuations (laser linewidths) much larger than the Schawlow-Townes limit~\cite{SchalowTownes58} and are subject to
continuous \mbox{long-term} frequency drifts. For example, free- \linebreak running diode lasers show a laser linewidth of about 100~MHz and frequency
drifts of up to many 10~MHz/h, due to temperature and air pressure fluctuations in the laboratory, diode current instabilities etc.~\cite{Wieman1991}.
Whereas the linewidth of a diode laser can be narrowed to typically less than 1~MHz using the passive optical feedback from a
grating~\cite{Wieman1991,Ricci1995}, the unwanted frequency drifts on longer time scales are more difficult to eliminate. A common technique is to
actively stabilize the laser onto an external high-finesse Fabry-Perot interferometer (FPI) or, alternatively, onto suitable atomic or molecular
transitions. For the latter, transition lines of various substances, e.g., alkaline atoms or molecules like iodine or tellurium, have been widely in use.



The thorough investigation of the iodine spectrum by Gerstenkorn et al. \cite{Gerstenkorn1,Gerstenkorn2} together with the ease with which an iodine
vapor cell can be handled make iodine a particularly attractive element for this purpose \cite{Nevsky2001,Picard2003,Hong2004,Skvortsov2004}.
On the other hand there are also a number of drawbacks: the main spectrum of iodine ranges from 500 to 900~nm, and within this range the number
of strong lines is limited so that for many frequencies of interest no excitation lines for the laser frequency stabilization are available.
In this case it is principally always possible to use an external FPI to stabilize the laser. However, to set up an external \mbox{frequency-stable} FPI may entail considerable experimental efforts and, moreover, cannot completely eliminate the unwanted residual frequency drifts. Therefore, if possible, it is usually advantageous to use atomic or molecular transitions as a reference for the active laser frequency stabilization since the atomic excitation frequencies do not change in time. Apart from iodine, there is thus a high demand to search for other atoms or molecules suitable for this purpose.

In this paper, we investigate a hitherto unexplored element in use for active laser frequency stabilization, namely the atom
pra\-seo\-dym\-i\-um. Because of its five outer valence electrons, the energy spectrum of pra\-seo\-dym\-i\-um exhibits an extremely rich level structure starting already $4\,432.22 \ \mathrm{cm^{-1}}$ above the ground state.
Since the spectrum resembles more that of a solid than that of an atom, it is sometimes also called `quasi-continuous'. The pra\-seo\-dym\-i\-um spectrum has been well investigated by several groups over the last few decades \cite{Ginibre0,Ginibre1,Guthoehrlein,Bakkali,Windholz,Zalubas1973,Furmann1997,Furmann2006,Martin1978,Childs1981,Kuwamoto1996,Boeklen1975,Macfarlane1982,Furmann1998,Krzykowski1998}. So far, more than 1200 energy levels and \mbox{25$\,$000} transition lines are known in the spectral range from 320~nm to 3.5~$\mathrm{\mu}$m of which one third has been classified so far. This multitude of excitation lines, including a rich hyperfine structure (hfs) for each resonance, in combination with the comprehensive knowledge of the spectrum make praseodymium an attractive element for the frequency stabilization of laser sources.

In what follows, we report the first active frequency stabilization of a laser onto a praseodymium resonance. Several praseodymium transitions between
1105~nm and \linebreak 1123~nm were systematically investigated for this purpose, corresponding to the tuning range of the diode laser used in the
experiment. The transitions were excited in a hollow cathode lamp (HCL) and recorded via laser-induced fluorescence (LIF). Active frequency
stabilization of the diode laser was demonstrated using standard laser locking techniques. In this way, the \mbox{long-term} frequency drift of the
diode laser of more than 30~MHz/h was eliminated and its frequency continuously stabilized to within 1.4(1)~MHz. In addition, one new strong
praseodymium transition close to the 4$^{th}$~sub-harmonic of the \Dtwo~transition \MgIonLine~of \linebreak \MgIon~at \mbox{$\lambda_{Mg^{+}}^{4^{th}SH}$~=~4~$\cdot$~$\lambda_{Mg^+}$~=~1118.540~nm} was found, making it possible to actively lock the diode laser at this wavelength. After frequency quadruplication this will allow the continuous excitation of the \MgIon~\Dtwo~transition at \MgLine.

The paper is organized as follows: In Sects. \ref{Pr} and \ref{SetupLIF} we present the element praseodymium and introduce the experimental setup used to obtain the praseodymium LIF spectra. In Sects. \ref{LIF} and \ref{Lock} we display several experimental LIF spectra recorded in the range 1105 to 1123 nm and discuss the possibility to actively frequency stabilize a laser onto one of these lines. In Sect. \ref{Mg} we present a new praseodymium transition, close to the 4$^{th}$~sub-harmonic of the \Dtwo~line of \MgIon. Frequency quadruplication of an infrared laser, locked to this transition, would make it possible to continuously excite the \Dtwo~transition in \MgIon. In Sect. \ref{conclusion} we finally draw conclusions.

\section{The element praseodymium}
\label{Pr}

The chemical element praseodymium (Pr) belongs to the lanthanide series (rare earths group) and is placed between the elements cerium (Ce) and neodymium (Nd). One of the prominent Pr electron configurations with three valence electrons inside the $4f$ shell is [Xe]$4f^3\,6s^2$. The lowest term under a total of 17 terms of this configuration is a $^4I$, where \PrTerm \ corresponts to the electronic ground state of Pr. Hereby, the designation \PrTerm~describes the coupling of the three equivalent electrons in the open $4f$ shell in LS-notation (LS-coupling nomenclature: $^{2S+1}L^{P}_{J};\ P$ denotes the parity of the level and $J=L-S$ in this case).

The naturally occurring Pr is composed of only one stable isotope \hspace{6pt} \element{Pr}{ 59}{141} with nuclear spin $I=5/2$.
The hyperfine interaction causes a hyperfine level splitting determined by the magnetic dipole coupling constant $A$ and the
electric quadrupole coupling constant~$B$ (magnetic dipole moment $\mu_I$ = 4.2754(5)~$\mu_N$ \cite{Macfarlane1982}, electric quad\-ru\-pole moment $Q$
= -0.0024~b \cite{Boeklen1975}). For a hyperfine sublevel characterized by the quantum number $F$ the shift $\Delta E_{h\!f\!s}$ with respect to the
corresponding fine structure level energy can be written in first order perturbation theory in the form

\small
\begin{equation}
\label{hfs}
\Delta E_{h\!f\!s} = \frac{1}{2} K A + \frac{\frac{3}{4}K(K + 1) - I(I + 1)J(J + 1)}{2I(2I - 1)J(2J - 1)} B
\end{equation}
\normalsize
with
\small
\begin{eqnarray}
\label{K}
K = F(F + 1)-I(I + 1)-J(J + 1),
\end{eqnarray}
\normalsize
where $I$, $J$, and $F$ are the nuclear spin, the total electron and total atom angular momentum quantum numbers, respectively. Eq.(\ref{hfs}) is
usually called \textit{Casimir's formula}. Since the Pr nucleus is almost spherical, the electric quadrupole term can be largely neglected in Eq.(\ref{hfs}).

The hyperfine splitting leads to a manifold of hfs sublevels. The number of sublevels is given by:
\begin{eqnarray}
\label{J_greater_I}
& 2J + 1 & \ \ \mathrm{for} \ \ \, J < I \ \ \, \mathrm{or} \\
& 2I + 1 & \ \ \mathrm{for} \ \ \, J \geq I
\end{eqnarray}
If at least one of the $J$-values of the combining fine structure levels is greater than 5/2, then the hyperfine structure shows a flag pattern with six
main (diagonal) hyperfine components \mbox{($\Delta F = \Delta J$)}. Besides these main components two additional groups of weaker off-diagonal
hyperfine components \mbox{($\Delta F \neq \Delta J$)} can be observed where the number of off-diagonal elements depends on $\Delta J$. If $\Delta J =
0$ one can find 10 off-diagonal components whereas if $\Delta J = \pm 1$ there are only 9 off-diagonal components can be observed. Since the frequency spacing between neighboring hyperfine components is typically on the order of 2~GHz and with Doppler-widths of approximately 400~MHz at 1115~nm the
full spectrum of the hfs can be usually resolved. To obtain the spectra the spectroscopic method of LIF is employed using a specially designed HCL.

The active laser frequency stabilization onto a particular Pr transition can be roughly divided into two parts. The first part deals with the
generation of a strong LIF signal in the HCL with high signal-to-noise (S/N) ratio.
The second part refers to the laser frequency stabilization itself, i.e., the stabilization of the laser onto a particular LIF resonance. Both parts
will be discussed in more detail in the following three sections.

\begin{figure}[t!]
 \centering
 \includegraphics[width=0.40\textwidth]{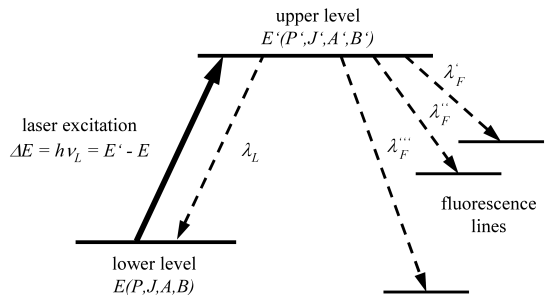}
 \caption{Part of the level scheme used for the active frequency stabilization of a diode laser via LIF detection. Resonant laser excitation at the wavelength $\lambda_{L}$ increases the population of the upper level $E'$. Therefore the fluorescence intensity of the light stemming from the spontaneous decay of this level is modified. This modification is recorded and used for the active laser frequency stabilization. The fluorescence light is usually composed of a set of different fluorescence channels with wavelengths $\lambda'_F$, $\lambda''_F$, $\lambda'''_F$, \ldots, which give information on the excited state. Each energy level is characterized by four parameters, namely $P$, $J$, $A$, and $B$. This set of parameters represents an atomic 'fingerprint' of an energy level and can be calculated from the observed LIF spectrum. This information in connection with the set of measured LIF wavelengths allows to classify the transition.}
 \label{fig:Prinzip_LIF}
\end{figure}

\section{Experimental setup for generating LIF spectra}
\label{SetupLIF} 

LIF spectroscopy in combination with an HCL is a sensitive and \mbox{widely-used} technique of recording the spectra of various gaseous substances. The
broad-band excitation in the hollow cathode discharge leads to a highly-excited plasma with a local thermodynamic equilibrium of the population
distribution among the atomic energy levels. Therefore, LIF in an HCL makes it also possible to investigate the level schemes of high-lying energy states, in the case of Pr of up to \mbox{$30\,000 \ \mathrm{cm^{-1}}$} above the ground state.

\begin{figure}[t!]
 \centering
 \includegraphics[width=0.42\textwidth]{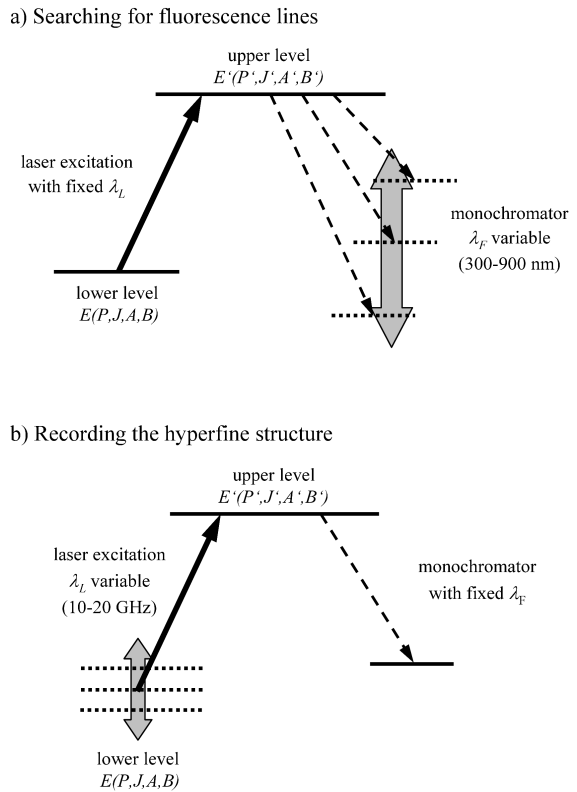}
 \caption{Two kinds of LIF experiments can be performed with our experimental setup: a) For the search of new fluorescence lines the laser frequency is fixed to a resonance and the transmission wavelength of the monochromator is varied. b)~For the precise recording of the hyperfine structure of the selected transition the monochromator is fixed to a strong LIF line and the laser frequency is varied.}
 \label{fig:Prinzip_LIF_2Arten}
\end{figure}

The basic principle of LIF spectroscopy is depicted in Fig.~\ref{fig:Prinzip_LIF}. The laser beam with wavelength $\lambda_L$ is led through the plasma
column of the hollow cathode discharge and is slowly scanned over the spectral range of interest. The selective absorption of the laser beam increases
the population of the upper level $E'$ when hitting a resonance. This change in the population distribution can be detected by measuring the rise of
fluorescence light intensity at the wavelengths $\lambda'_F, \lambda''_F, \lambda'''_F, \ \ldots$, emitted by the spontaneous decay from the upper
level $E'$ to corresponding lower levels (LIF). A LIF signal will only be recorded if the laser frequency is in resonance with a dipole allowed transition between certain hyperfine sublevels $F$ of the lower level $E$ and $F'$ of the upper level $E'$. A tunable monochromator (spectral bandwidth $\Delta\lambda$~=~0.2~nm) which registers the fluorescence light makes it possible to separate all decay channels of the fluorescence spectrum. At the same time, the wavelength band-pass filter of the monochromator avoids an overload of the photomultiplier tube caused by the bright light emitted by the hollow cathode plasma over the whole spectral region.

There are two conceptionally different kinds of LIF experiments, depending on the operation of the mo\-no\-chro\-ma\-tor. For the first type of
experiments, the laser frequency is tuned to an optical resonance, so that the monochromator can be used to analyze the LIF emission spectrum. Since every energy level decays spontaneously into particular final states depending on the selection rules, the monitored fluorescence light can give
information on the position, parity and quantum number $J$ of the upper state.
In the other type of experiment, the monochromator is tuned to a strong LIF line and the laser frequency is then scanned over a frequency region of
interest to obtain the hyperfine structure of the excitation line. In this way, the entire hfs of a spectral line can be studied. Both LIF methods are
displayed in Fig.~\ref{fig:Prinzip_LIF_2Arten}. The last mentioned method has been applied mostly in this work.

\begin{figure}[t!]
 \centering
 \includegraphics[width=0.47\textwidth]{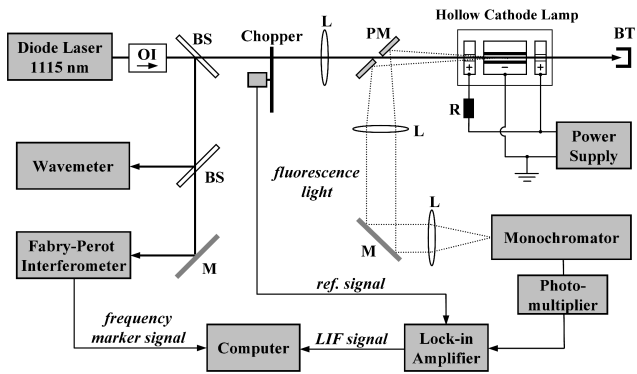}
 \caption{Experimental setup for LIF spectroscopy in an HCL. The terms OI, BS, L, M, MP, and BT stand for optical isolator, beam splitter, lense, mirror, mirror with pinhole, and beam trap, respectively. For a detailed description of the experimental setup see text.}
 \label{fig:setup_LIF}
\end{figure}

The experimental setup for the generation of the LIF spectra is presented in Fig. \ref{fig:setup_LIF}. A commercial diode laser from Toptica Photonics (DL 100) at 1115~nm center wavelength is used for the excitation of Pr atoms produced by sputtering inside the HCL. The grating-stabilized extended cavity diode laser (ECDL) has a linewidth of less than 1~MHz which is sufficiently narrow to record the Doppler-broadened Pr resonances. The ECDL has a maximum output power of 83~mW at \linebreak 208~mA and is tunable over a range of almost 20~nm, from 1105 to 1123 nm; without mode hopping the laser can be tuned in single mode over 20~GHz. The internal collimator lens of the ECDL is used to focus the laser beam into the HCL to yield optimum illumination of the reaction volume. In order to prevent unwanted optical feedback from the HCL and other optical elements an optical isolator with isolation $>$~60~dB is installed in front of the ECDL.

A beam splitter placed behind the optical isolator feeds parts of the laser light into a wavemeter and an FPI. Both devices are used to determine the
laser frequency. The wavemeter (uncertainty~50~MHz) is used for the absolute frequency calibration of the LIF spectra whereas the FPI (free spectral
range 2.109(12)~GHz) serves as an additional frequency marker in order to linearize the recorded LIF spectra.

\begin{figure}[t!]
 \centering
 \includegraphics[width=0.485\textwidth]{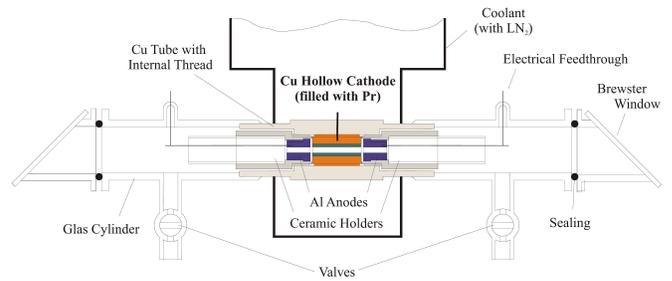}
 \caption{(Color online) HCL in cross-section. For details see text.}
 \label{fig:HCL}
\end{figure}

The non-commercial HCL especially designed for this experiment consists of several parts (see Fig.~\ref{fig:HCL}). The central element of the
symmetrical HCL is a copper tube into which an axially hollowed cathode of 20 mm length is screwed, filled on the inner walls with Pr. In addition,
axially hollowed anodes made from aluminum are fixed at both ends of the copper tube. The inner diameter of the hollow cathode, best for the
investigation of neutral Pr, has been found to be 3~mm. The copper tube together with the cathode and the two anodes is placed inside a glass cylinder,
sealed at both ends with Brewster windows to allow maximum transmission of the laser beam through the HCL.

The HCL was operated with a buffer gas (argon) at a pressure of 0.5~mbar. In order to obtain a constant discharge current of about 60~mA, a voltage of 300~V was applied. Under these conditions one can assume that the fraction of atomic Pr in the plasma amounts to more than 90~percent, the rest of the
sputtered material being mainly singly ionized Pr$^+$.

In order to achieve a more stable discharge, the HCL can be cooled with liquid nitrogen. The nitrogen cooling results in an internal temperature of the
HCL of about 100 to 200~$^\circ$C (Boltzmann temperature) which greatly reduces the Doppler-widths of the spectra down to typically 350 to 400~MHz at
1115~nm.

Passing laser radiation through the gas discharge \linebreak yields a new population distribution among the two energy levels of Pr participating in a resonance. The fluorescence light that leaves the HCL in case of resonant laser excitation is collected by a pin-hole mirror (placed on the side of the incoming laser beam) and is directed to the entrance slit of the monochromator by a set of mirrors and lenses. The monochromator with a tuning range from 300~to~900~nm and a resolution of $\Delta \lambda / \lambda \approx 3 \cdot 10^{-4}$ can be used to filter single LIF lines out of the total fluorescence spectrum. A photomultiplier with high sensitivity for the visible spectrum is mounted at the exit slit of the monochromator allowing the detection of extremely low fluorescence signal changes. In addition, a mechanical chopper (280~Hz) placed inside the laser beam enables a phase-sensitive detection and amplification of the LIF signals by use of a lock-in amplifier (LIA). Even very low-intensity LIF signals in an extremely noisy environment can thus be observed and recorded.

\begin{figure}[t!]
 \centering
 \includegraphics[width=0.48\textwidth]{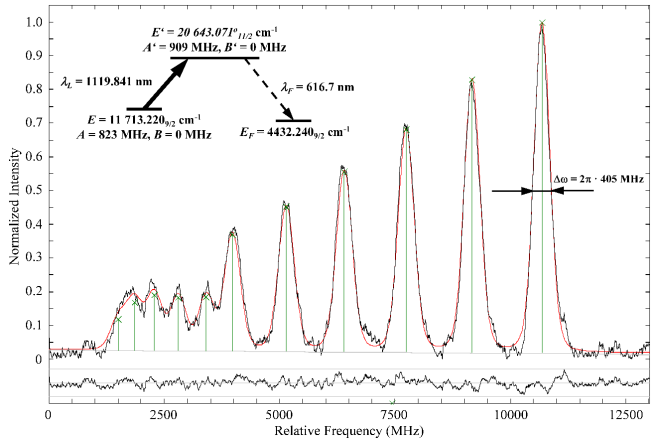}
 \caption{(Color online) Hyperfine structure of the spectral line $\lambda_{L}$~=~1119.841~nm of neutral Pr recorded in an HCL via LIF. $\lambda_L$ is the vacuum wavelength of the laser excitation defining the center position of the hyperfine structure of the transition. After laser excitation of a transition between the hyperfine structure sublevels of the lower level $E = 11\,713.220_{9/2} \ \mathrm{cm^{-1}}$ and the upper level $E' = $ \textit{20\,643.071}$^o_{11/2} \ \mathrm{cm^{-1}}$, according to the selection rules of electric dipole radiation, the excited sublevels of the upper level $E'$ decay into the energy level $E_F = 4\,432.240_{9/2} \ \mathrm{cm^{-1}}$ by emission of the fluorescence light at $\lambda_{F}$ = 616.7 nm (given in air). The Voigt-width of the hyperfine components is 405~MHz. The figure shows the experimental excitation spectrum and the corresponding theoretical fit on a relative frequency scale. The theoretical profile was calculated by a especially designed evaluation program \cite{Fitter} based on a least-squares-fit method. The plotted line below the spectrum is called the deviation curve and displays the difference between the theoretical fit and the experimental curve. It illustrates the quality of the fit. This Pr spectral line was newly classified during this work.}
 \label{fig:bestfit}
\end{figure}

The LIF signal together with the marker signal from the FPI are simultaneously recorded and stored in a computer. The acquired LIF spectra
can then be processed by an especially designed evaluation program \cite{Fitter}. This program serves to determine the hfs parameters of the measured
lines by fitting the experimentally recorded spectra by a theoretical curve composed of a specified number of Voigt profiles using the method of
least- \linebreak squares. Fitting parameters of the theoretical curve are the hyperfine constants $A$ and $B$ of the involved lower and upper energy levels, the linewidths of the Gaussian and Lorentzian parts of the Voigt profiles, and the relative intensities of all hyperfine components. The total angular momentum quantum numbers $J$ and $J'$ of the lower and upper levels can be chosen separately before the algorithm of the fit is carried out. The choice of specific $J$ values defines the number of hyperfine components and is therefore a prerequisite for a good theoretical reconstruction of the recorded hfs patterns. As the positions of a maximum of 16 (if $\Delta J = 0$) or 15 (if $\Delta J = \pm 1$) hfs components obey the Casimir relations (see Eq.(\ref{hfs})), only one additional parameter (the center of gravity position of the line) is needed. The final fit and the experimental spectrum are plotted together. An additional curve called deviation curve is shown at the bottom of each spectrum. This curve displays the error between the best fitted theoretical and the experimental curves and illustrates the quality of the fit. Fig.~\ref{fig:bestfit} shows an example of a typical LIF spectrum together with the adjusted theoretical curve in the best-fit situation.

\begin{figure}[t!]
 \centering
 \includegraphics[width=0.48\textwidth]{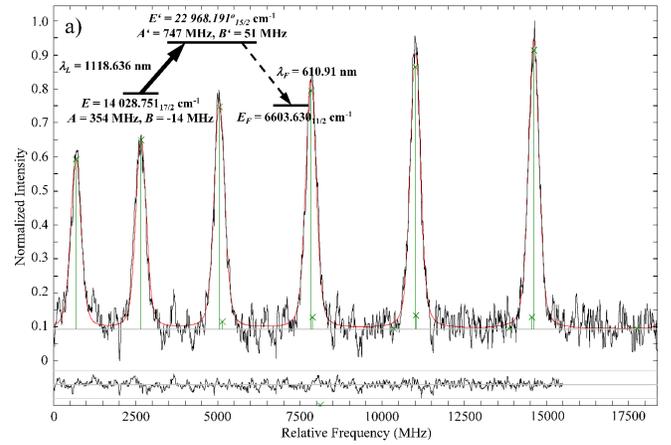} \\ \vspace{5mm}
 \includegraphics[width=0.48\textwidth]{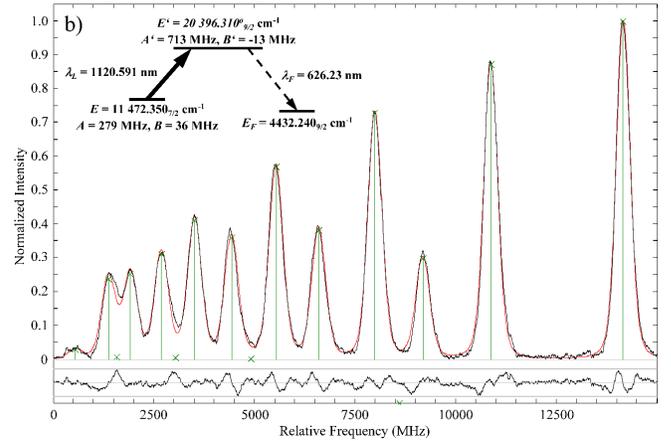} \\ \vspace{5mm}
 \includegraphics[width=0.48\textwidth]{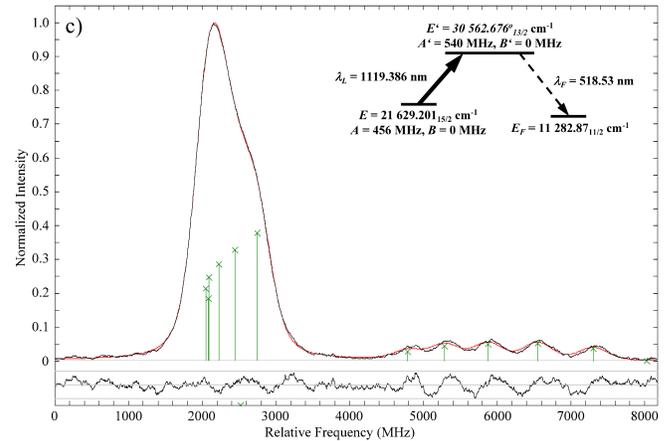}
 \caption{(Color online) Three typical hyperfine structures of neutral Pr spectral lines recorded in an HCL via LIF: (a)~Confirmation of an already classified Pr line by Ginibre \cite{GinibreGuthoehrlein}; (b) and (c) Transitions already found by Ginibre, but unclassified so far. 
 For further de\-tails see text.}
 \label{fig:Spektrum_S12}
\end{figure}

\section{LIF spectroscopy between 1105 and 1123~nm}
\label{LIF}
In the first stage of our investigations of neutral Pr we recorded a number of lines in the range between 1105 and 1123~nm, accessible with the tunable
ECDL. Since up to now all LIF experiments with Pr have been performed using Ti:Sapphire or single-mode dye lasers, the Pr spectra recorded so far have
been limited to laser excitation in the visible range of the spectrum; apart from experiments employing Fourier spectroscopy \cite{Ginibre0,Ginibre1}
no direct electric dipole transitions in the infrared range of the spectrum have thus yet been observed.

The spectral line density and the amount of known lines in the infrared region between 1105 and 1123~nm is almost ten times lower compared to the
amount of lines classified in the visible. 
Since LIF spectroscopy is based on an optical coincidence measurement, where simultaneously the laser excitation and the detection window of the
monochromator must coincide with a dipole allowed excitation line and a radiating emission line, respectively, it is a hopeless undertaking to find a
resonance line in this wavelength region just by trial and error. (The wavelength range between 1105 and 1123~nm corresponds to $\approx$~4~THz; if the linewidth of a hfs component is $\approx$~400~MHz, the possibility of hitting the transition will be 1~:~10~000. In combination with the monochromator tunable in the range 300 to 900~nm and a bandwidth of $\Delta\lambda$~$\approx$~0.2~nm the overall possibility of recording a LIF signal is in the best case 1~:~30~000~000.)

To facilitate the task, we made use of a special data\-base in which all fluorescence lines as well as all energy levels of Pr recorded so far
are collected \cite{Ginibre0,Ginibre1,Guthoehrlein,Bakkali,Windholz}. The database contains not only the almost 600 levels of neutral Pr given in Ginibre's published and unpublished work \cite{Ginibre0,Ginibre1,GinibreGuthoehrlein} but also about 600 further levels, most of them discovered in Hamburg \cite{Guthoehrlein,Bakkali}. The constantly updated database gives information on the wavelengths of all recorded excitation lines and their combining energy levels, including $J$ quantum numbers, hyperfine coupling constants $A$ and $B$, and parities $P$. These characteristic data are supplemented by a set of fluorescence wavelengths which were observed in earlier LIF experiments \cite{Guthoehrlein,Windholz,Bakkali,Furmann1997,Furmann2006,Childs1981,Kuwamoto1996,Furmann1998,Krzykowski1998}. Apart from the identification of already measured lines and transitions the database allows in particular the prediction of new spectral lines and energy levels as well as the classification of hitherto unidentified hyperfine spectra.

According to this database, 66 lines were measured between 1105 and 1123~nm, which corresponds to the tuning range of the diode laser. Only a few of them had been classified so far. For all other lines only the transition wavelengths and no further information about the combining energy levels were listed.
Since it was not our intention to check all 66 measured lines, we picked out a few of them, in particular \mbox{low-lying} combining energy levels. \mbox{Low-lying} energy levels are usually much more populated than the higher energy levels, so that the corresponding LIF resonances are expected to be much stronger. This property becomes all the more important, as the laser power of the ECDL was restricted to 60~mW inside the HCL.


In Fig.~\ref{fig:Spektrum_S12} three typical hfs spectra with different S/N ratios around 1119~nm are shown, together with the corresponding
excitation and emission schemes. The wavelengths of the fluorescence lines are given in air. Spectrum (a) is a confirmation of a classification given in \cite{GinibreGuthoehrlein}. Spectra (b) and (c) were observed on the base of new levels taken from \cite{Guthoehrlein}.

\begin{figure}[t!]
 \centering
 \includegraphics[width=0.47\textwidth]{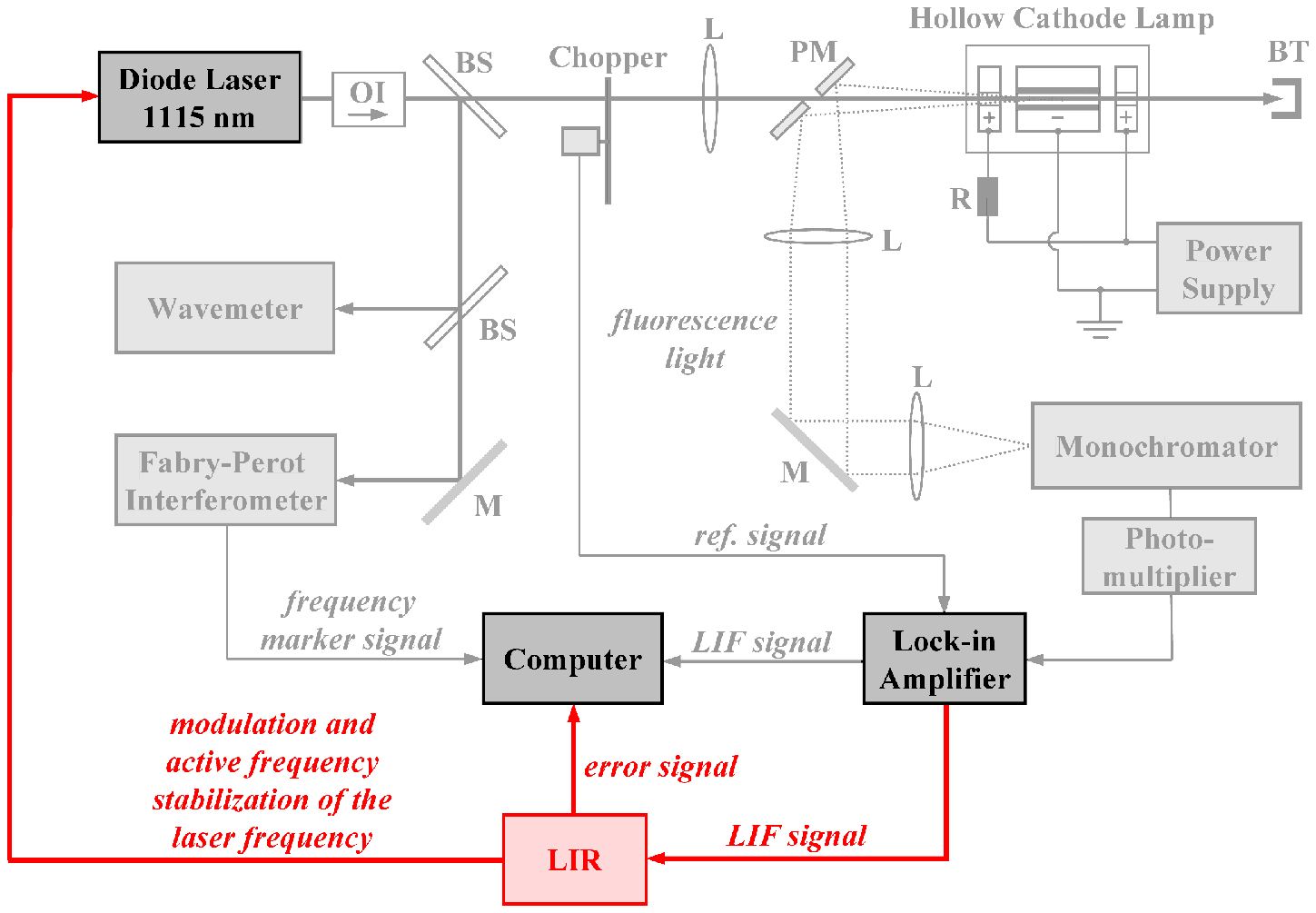}
 \caption{(Color online) Experimental setup for the active frequency stabilization of the diode laser onto a particular hyperfine structure component of Pr. The LIR generates a dispersive signal (first derivative of the LIF signal) which serves as the error signal for the PID to actively stabilize the laser frequency onto the line center. For more details see text.}
 \label{fig:setup_LIF_Lock}
\end{figure}

The particular hfs patterns originate from the different types of electronic transitions, defined by \linebreak $\Delta J = 0, \pm 1$, as well as from the hyperfine splitting of the combining levels: well separated main hyperfine components with a characteristic flag pattern can be explained by clearly different hyperfine splitting of the lower and upper energy levels. This is the case for example for the spectra shown in Fig.~\ref{fig:Spektrum_S12} (a) and (b). On the contrary, the spectrum in Fig.~\ref{fig:Spektrum_S12} (c), apart from five small off-diagonal components, displays only one broad asymmetric structure, resulting from the overlap of all six main components because of almost equal splitting of the lower and upper energy levels. From all three spectra shown in Fig.~\ref{fig:Spektrum_S12} the spectrum in Fig.~\ref{fig:Spektrum_S12} (b) offers the best properties for an active frequency stabilization of a laser due to the good S/N ratio and the clearly separated hfs components.


\begin{figure*}[t!]
 \centering
 \includegraphics[width=0.8\textwidth]{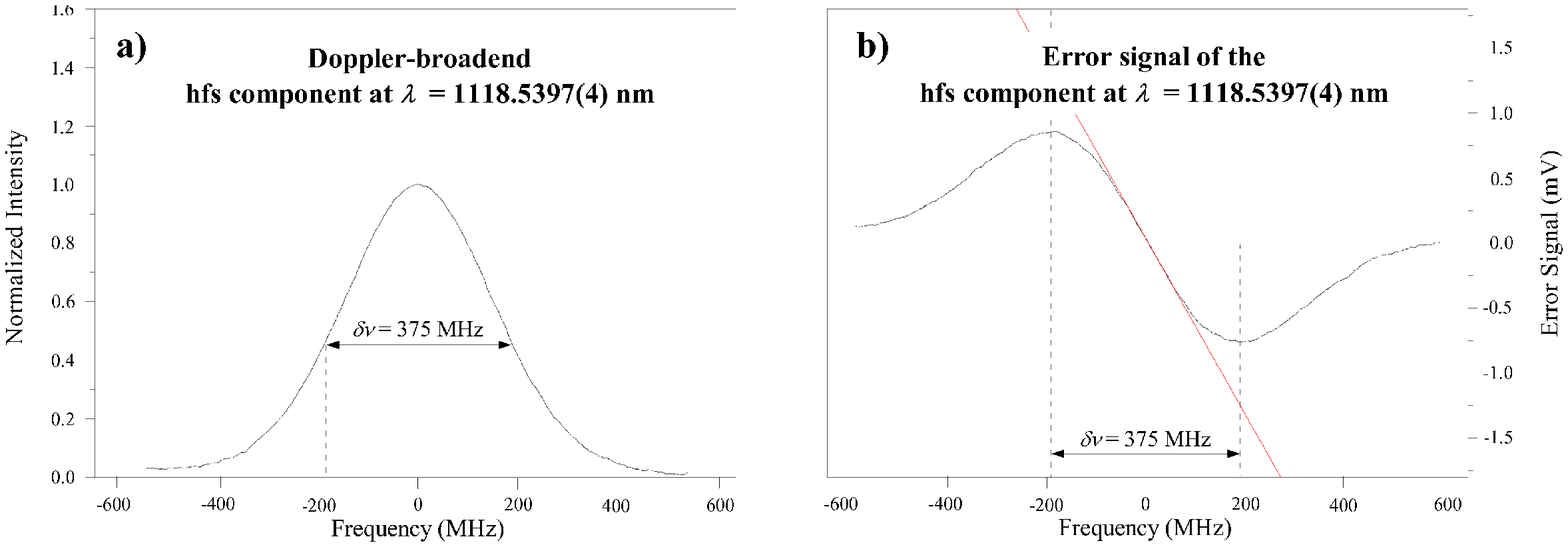} \\ \vspace{5mm}
 \includegraphics[width=0.8\textwidth]{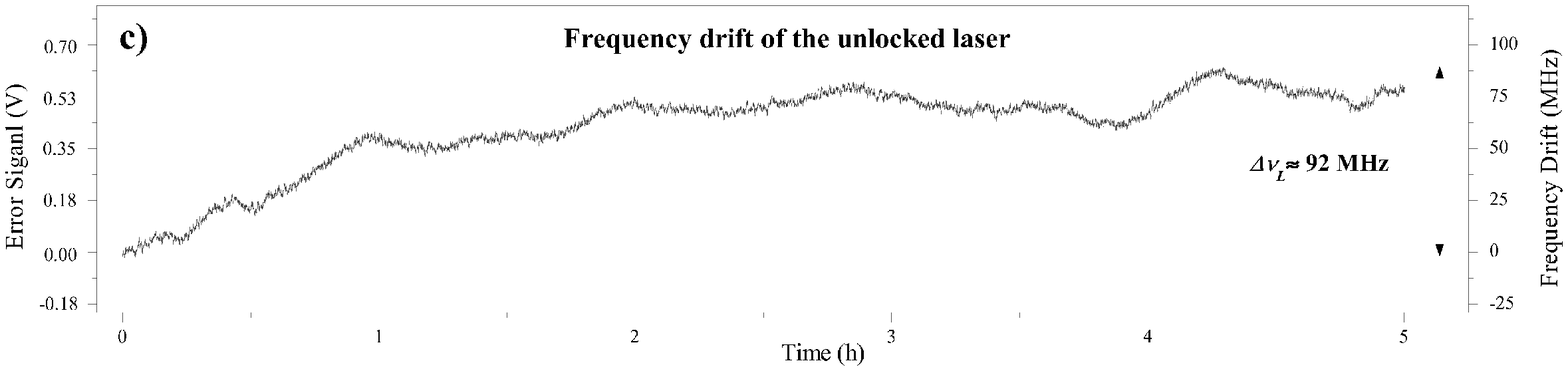}
 \caption{a) LIF signal of the strongest hyperfine structure component \mbox{$F = 6 \rightarrow F' = 7$} of the newly classified Pr transition \PrLockLine \ and b) the corresponding dispersive error signal generated by the LIR module. c) Frequency drift $\Delta\nu_L$ of the \mbox{free-running} diode laser.}
 \label{fig:drift}
\end{figure*}

\begin{figure*}[t!]
 \centering
 \includegraphics[width=0.8\textwidth]{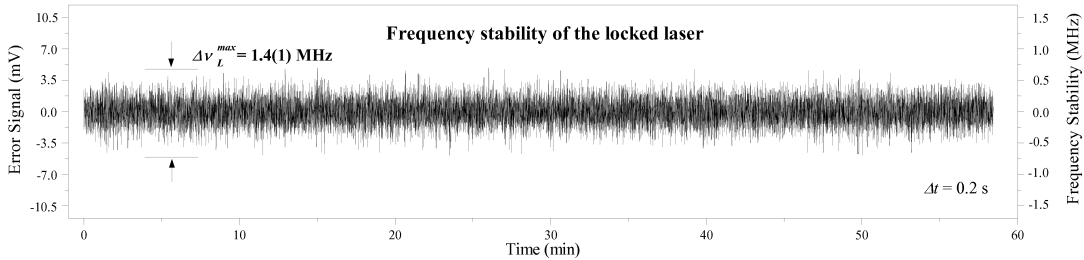}
 \caption{Long-term frequency stability of the infrared diode laser at \PrComponent~after locking the laser onto the maximum of the strongest hyperfine structure component of the Pr transition \PrLockLine. The laser frequency for averaging times $\Delta t \geq 0.2$~s remained within an absolute bandwidth of $\Delta\nu_L^{max}$ = 1.4(1)~MHz.}
 \label{fig:Lock_Pr_Lock}
\end{figure*}

In the case that no resonance lines are known in a particular wavelength region, the already mentioned database for the lines and energy levels of Pr
together with a software developed at the Helmut-Schmidt- \linebreak Universit\"at Hamburg, Germany, makes it possible to \linebreak search for new transitions. (Note that L. Windholz's group from the TU Graz, Austria, developed a similar program. It is free to the public and can be used for any element.) The list of theoretically proposed transitions contains hereby not only the information on the wavelength needed for the laser excitation and their combined energy levels but also the known fluorescence lines of the levels. The latter can be used for the identification of the laser-induced transition. The question remains, however, whether the theoretically predicted transitions can be actually confirmed in the experiment.

Since the amount of known resonance lines in the infrared wavelength range is relatively low, the program turned out to be especially helpful. For example, the program predicted ten transitions between $\lambda_a$ = 1118.519~nm and $\lambda_b$ = 1118.563~nm ($\Delta\nu \approx 11 \ \mathrm{GHz}$) close to the 4$^{th}$~sub-harmonic of the \Dtwo~transition of \MgIon~at \MgLineSH which have never been excited or recorded before. However, in most of these resonances the transitions start from relatively \mbox{high-lying} energy levels, $15\,000 \ \mathrm{cm^{-1}}$ to  $23\,000 \ \mathrm{cm^{-1}}$ above the ground state, for which the chance of recording a strong LIF signal is low.

With the available laser power of 60~mW inside \linebreak the HCL only two of the ten predicted Pr lines were \linebreak confirmed in the experiment, namely those with \linebreak the lowest initial states ($E_1 = 14\,981.51 \ \mathrm{cm^{-1}}$ and \linebreak $E_2 = 16\,502.61 \ \mathrm{cm^{-1}}$). They were detected by their strongest fluorescence decay channels at 567.75 and \linebreak 573.58~nm, respectively. The latter Pr line is shown in Fig.~\ref{fig:HFS_Lock}. Both lines were recorded and classified for the first time. The two lines show a flag pattern with clearly separated hfs components and linewidths of about 375~MHz. With a S/N of about 20 and 70, respectively, it is relatively simple to stabilize a laser onto any of the six main hfs components of each of the two lines. Moreover, the presented S/N ratios could be improved by using a higher laser power. However, for the diode laser at 1115~nm, higher intensities can only be achieved with special amplifier units, e.g., tapered or fiber amplifiers. A larger laser intensity would most probably allow the detection of the other eight theoretically predicted Pr lines.



\begin{figure}[t!]
 \centering
 \includegraphics[width=0.47\textwidth]{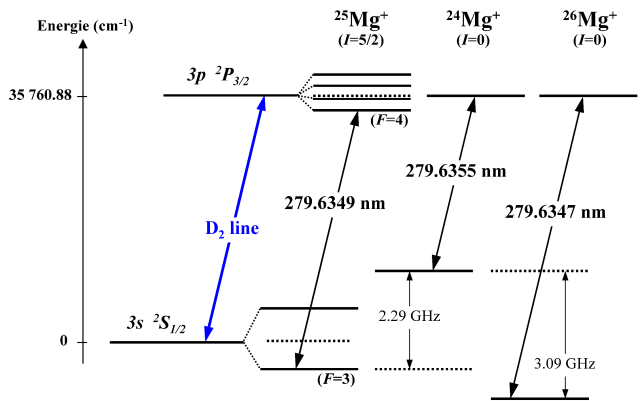}
 \caption{(Color online) Energy level scheme of the \Dtwo~line of singly ionized magnesium (\MgIon) with isotope shifts for the three naturally occurring isotopes \MgIons where the level isotope shifts of the $^2P_{3/2}$ states are set to zero \cite{Batteiger2009}.}
 \label{fig:Mg_level_scheme}
\end{figure}

\section{Laser stabilization using Pr transitions}
\label{Lock}

The active frequency stabilization of the infrared diode laser onto a particular Pr hfs component was performed using a so-called Lock-In-Regulator (LIR) from Toptica Photonics (see Fig.~\ref{fig:setup_LIF_Lock}). The LIR module is a phase- \linebreak sensitive lock-in detection unit with an integrated proportional, integral, and differential (PID) regulator. \linebreak Based on a modulation of the laser frequency by dithering the piezo of the ECDL grating the first derivative of the LIF resonance is generated. This error signal is used to accomplish via the PID regulator the active frequency stabilization of the laser or, outside the lock, to investigate the frequency fluctuations and drifts of the \mbox{free-running} laser. For the latter, the linear part of the error signal around the zero crossing is used as a frequency discriminator (see Fig.~\ref{fig:drift} (a) and (b)).
Fig.~\ref{fig:drift} (c) shows the drift behavior of the unlocked laser. For the measurements the strongest hfs component of the Pr transition

\begin{center}
 \PrLockLine\\
 \hspace*{-2pt} $F = J + I = 6 \rightarrow F' = J' + I = 7$
\end{center}
at \PrComponent~was used, having a linewidth of 375 MHz (see Fig.~\ref{fig:HFS_Lock}). Within five hours of measurement a drift of more than 90~MHz was observed.

After active frequency stabilization of the laser this frequency drift was strongly reduced. Fig.~\ref{fig:Lock_Pr_Lock} shows the \mbox{long-term} frequency stability of the laser (averaging \linebreak time~$> 0.2$~s) after actively locking the laser onto the mentioned hfs component. As can be seen in the figure the laser frequency was stabilized to within 1.4(1)~MHz for an unlimited amount of time.

The ultimate frequency stability of the laser is determined by the S/N ratio of the LIF signal. Since the LIF signal is relatively low and furthermore obscured by the bright fluorescence background of the HCL, it is necessary to work with a lock-in amplifier (LIA) (see Sect.~\ref{SetupLIF}). The LIF signal, after amplification by the LIA, is fed into the LIR to produce the error signal for the active frequency lock (see Fig.~\ref{fig:setup_LIF_Lock}). Since the laser frequency is also modulated for the lock-in detection unit of the LIR, a too large time constant of the LIA actually improves the S/N of the LIF signal but at the same time also averages over the LIR modulation and thus suppresses the LIR error signal. A good compromise between a reasonable S/N of the LIF signal and an acceptable error signal of the LIR is therefore mandatory. For this purpose the modulation frequency of the LIR was reduced to the smallest selectable value (8.2~Hz). The bandwidth of the locking loop, given approximately by half of the modulation frequency of the LIR, therefore corresponded to about 4~Hz. This allows mainly the compensation of the low-frequency instabilities of the laser, i.e., the laser drift. 

\section{Using a Pr-stabilized laser for Mg$^+$ spectroscopy}
\label{Mg}

In the previous section we presented the active frequency stabilization of a diode laser onto a particular hyperfine transition of Pr at \PrComponent.
The selected Pr line served however merely as an example to demonstrate the feasibility of locking a laser onto a Pr resonance using LIF spectroscopy.
In principle, the diode laser could have been stabilized onto any other suitable Pr line in the range from 1105 to 1123~nm.

By frequency quadrupling the stabilized infrared \linebreak diode laser, \mbox{frequency-stable} \mbox{UV-light} in the range \linebreak 276 to 281~nm can be produced. This technique has already been employed to convert \mbox{frequency-stable} laser radiation at 922 nm to
narrow-band UV-light at 231~nm \cite{Schwedes2003}. Since diode lasers are not yet available at 280~or 560~nm, this approach makes it possible to transfer all well-known advantages of grating-stabilized ECDLs to these wavelength regions. This is an attractive alternative to (fre\-quen\-cy-doubled) dye lasers or tunable fiber lasers \cite{Friedenauer2006} otherwise employed in these wavelength regions. 

\begin{table}
\caption{Fundamental, 2$^{nd}$, and 4$^{th}$ sub-harmonics of the \mbox{\Dtwo~transitions} \MgIonLine~of the singly ionized magnesium isotopes \MgIons \cite{Batteiger2009}.}
\centering
\label{tab:Mg_lines}
\begin{tabular}{cccc}
\hline\noalign{\smallskip}
isotop & $\lambda$ (nm) & $\lambda^{2^{nd}SH}$ (nm) & $\lambda^{4^{th}SH}$ (nm)\\
\noalign{\smallskip}\hline\noalign{\smallskip}
$^{24}\mathrm{Mg}^+$ & 279.6355 & 559.2710 & 1118.5420\\
$^{25}\mathrm{Mg}^+$ & 279.6349 & 559.2698 & 1118.5396\\
$^{26}\mathrm{Mg}^+$ & 279.6347 & 559.2694 & 1118.5388\\
\noalign{\smallskip}\hline
\end{tabular}
\end{table}

As it turns out, this technique can also be used to produce \mbox{frequency-stable} laser light at 279.635~nm, making it possible to excite the \MgIonLine~\Dtwo~transition of singly ionized magnesium \MgIon. In recent years there has been a growing interest in \MgIon~since due to its internal level structure this ion is a favorite candidate for experiments in spectroscopy, quantum optics or in quantum information science \cite{Hoeffges1997,Schaetz2007,Friedenauer2008,Batteiger2009,Home2009,Schmitz2009}. The precise wavelengths of the \Dtwo~lines of the three stable
isotopes \linebreak \MgIons~together with the corre- \linebreak sponding line isotope shifts are shown in Fig.~\ref{fig:Mg_level_scheme}. \linebreak The corresponding fundamental, second, and fourth \linebreak sub-harmonics for the three isotopes are listed in \linebreak Table~\ref{tab:Mg_lines}.
\begin{figure}[t!]
 \centering
 \includegraphics[width=0.48\textwidth]{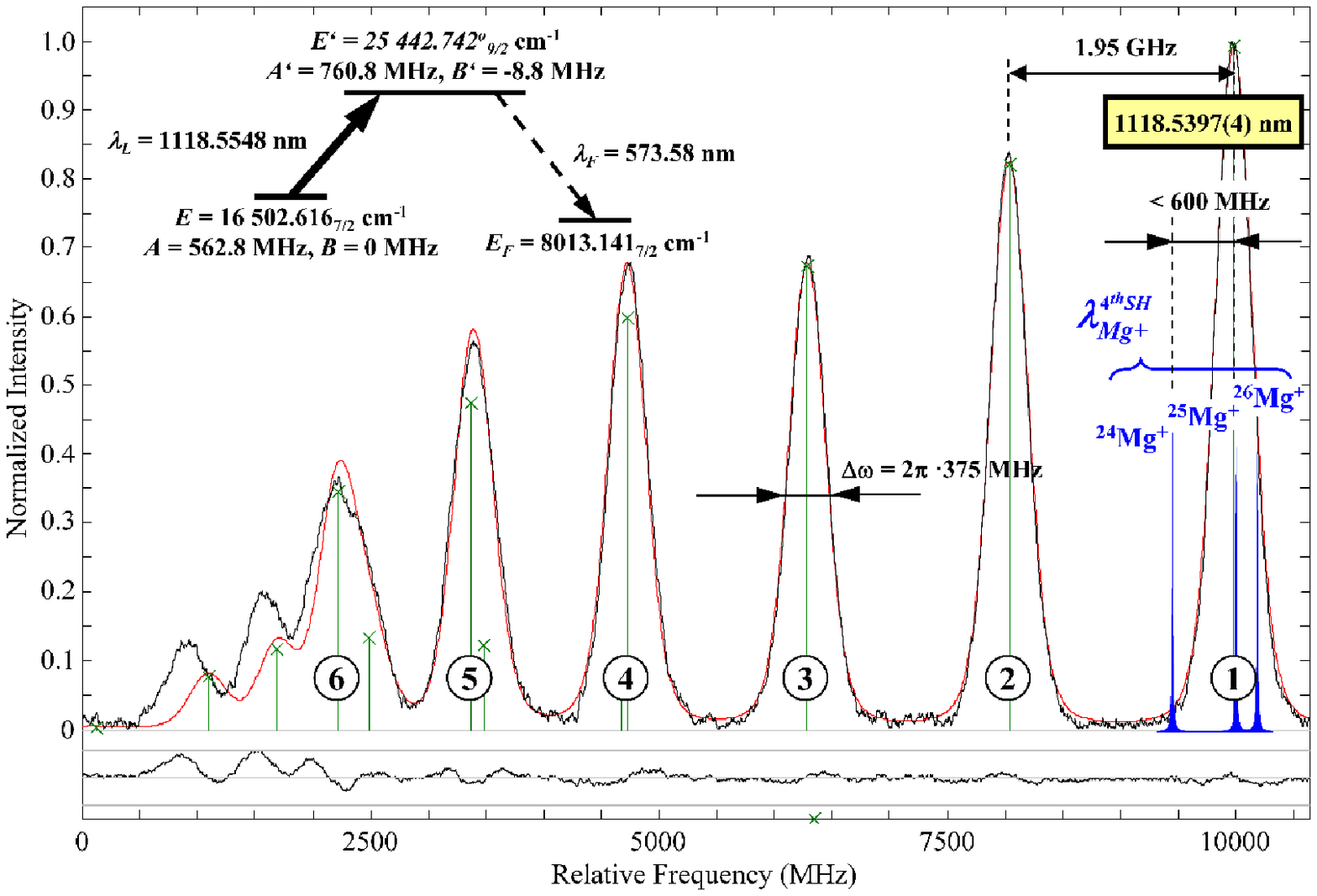}
 \caption{(Color online) Hyperfine structure of the Pr line $\lambda_L$~=~1118.5548~nm together with the 4$^{th}$ sub-harmonics~of the \Dtwo~transitions $\lambda_{Mg^{+}}^{4^{th}SH}$ of the three isotopes $^{24}$Mg$^+$, $^{25}$Mg$^+$, and $^{26}$Mg$^+$. The relative positions and the adapted linewidths \mbox{$\gamma_{Mg^{+}}^{4^{th}SH} = \gamma_{Mg^+} / 4 = 2 \pi \cdot 10.7$~MHz} are drawn to scale. Of the six components of the hyperfine structure, the strongest hyperfine component at \PrComponent~has a deviation of less than 600(100)~MHz with respect to $\lambda_{Mg^{+}}^{4^{th}SH}$~of the three isotopes and is therefore the closest and most suitable line for locking the frequency of the infrared diode laser in order to perform experiments with trapped \MgIon~ions.}
 \label{fig:HFS_Lock}
\end{figure}
The previously discussed strongest hfs com- \linebreak ponent \mbox{$F = 6 \rightarrow F' = 7$} of the Pr transition \linebreak \PrLockLine~at \PrComponent~is less than 600(100)~MHz away from the 4$^{th}$ sub-harmonics~of the \MgIon~\Dtwo~line $\lambda_{Mg^{+}}^{4^{th}SH}$ at 1118.540~nm~=~4~$\cdot$ $\lambda_{Mg^{+}}$ (see Fig.~\ref{fig:HFS_Lock}). This resonance can therefore be conveniently used to lock the laser in the infrared and - after frequency quadruplication - to obtain \mbox{frequency-stable} light in the UV in order to excite the \Dtwo~transition. The remaining frequency gap of 600(100)~MHz in the infrared can be easily bridged by an acousto-optical modulator (AOM). The second nearest hyperfine component $F  = 5 \rightarrow F' = 6$ of the transition \linebreak \PrLockLine~is \linebreak already almost 2 GHz away from $\lambda_{Mg^{+}}^{4^{th}SH}$, usually too large to be bridged by an AOM. Fig.~\ref{fig:HFS_Lock} shows the complete hyperfine structure of the transition \linebreak \PrLockLine~to\-geth\-er with $\lambda_{Mg^{+}}^{4^{th}SH}$~of the three stable isotopes $^{24}$Mg$^+$, $^{25}$Mg$^+$, and $^{26}$Mg$^+$. The relative positions and the line\-widths are drawn to scale. 
Other Pr hfs resonances investigated in this wavelength region exhibited either worse S/N ratios or were at least 5~GHz away from the desired $\lambda_{Mg^{+}}^{4^{th}SH}$.

The obtained \mbox{long-term} frequency stability of the laser of 1.4(1)~MHz at \PrComponent~- after frequency quadruplication - corresponds to a
frequency stability of $<$~6~MHz in the UV. This is by far sufficient for the resonant excitation of the \Dtwo~line of \MgIon~which has a natural
linewidth of \mbox{$\gamma_{Mg^+} = 2 \pi \cdot 43$}~MHz. By appropriately detuning the frequency-stabilized diode laser it thus becomes possible to excite or laser cool individual \MgIon~isotopes for an unlimited amount of time.

\section{Conclusion}
\label{conclusion}
In conclusion, we discussed the atom pra\-seo\-dym\-i\-um (Pr) as a versatile and hitherto unexplored reference for the active frequency stabilization of a laser. With five valence electrons Pr displays an extremely rich level structure, starting already $4\,432.22 \ \mathrm{cm^{-1}}$ above the ground state. So far, more than 1200 energy levels and \mbox{25$\,$000} transition lines between 320~nm and 3.5~$\mathrm{\mu}$m are known, including a rich hyperfine structure (hfs) for each resonance. This multitude of excitation lines together with the detailed knowledge of the spectrum makes Pr a very attractive reference for active laser frequency stabilization. To demonstrate the feasibility to use this element for the active frequency stabilization of lasers we actively locked a diode laser onto a particular Pr line. The corresponding excitation signals were recorded in a hollow cathode lamp (HCL) via laser-induced fluorescence (LIF) spectroscopy. Using standard locking techniques we were able to eliminate the frequency drifts of the unlocked diode laser of more than 30~MHz/h and stabilize its frequency permanently to within 1.4(1)~MHz for averaging times $>0.2$~s. The strongest hfs component of a newly found Pr line at \PrComponent~allows - after frequency quadruplication - the production of frequency stable UV light at \PrComponentQ~which can be used to excite the \Dtwo~transition in \MgIon. In future it might be possible to simplify the experimental design by replacing the HCL by a common rf-lamp filled with Pr.

%
%
\section*{Acknowledgement}
The authors would like to especially thank Prof. Hermann Harde from the Helmut-Schmidt-Universit\"at der Bundeswehr Hamburg for making measurements in his laboratory possible. The friendly assistance from Dr. Hans-Otto Behrens, G\"unther Helmrich, and Marion Kollmeier during the measurements in Hamburg was also appreciated. Thanks also go to Thomas Heine, Thorsten Schmitt, and Dr. Harald Ellmann from Toptica Photonics for their valuable technical support for the diode laser system. Furthermore, S.O. thanks the Universit\"at Bayern e.V. for financial support.


\end{document}